\pdfoutput=1
\documentclass{JINST}
\let\ifpdf\relax

\usepackage{upgreek}
\usepackage{float} % place figure where you like
\usepackage{amstext}
\usepackage{amsmath} 
\usepackage{amssymb}
\usepackage{amsfonts}
\usepackage{float} % place figure where you like
\usepackage{url}

\title{Performance Studies of a Micromegas Chamber in the ATLAS Environment}

\author{ Y. Kataoka$^a$, S. Leontsinis$^{b,c}$ and K. Ntekas$^c$\thanks{Corresponding
author.}~\\
\textbf{On behalf of the MAMMA collaboration}\\ 
\llap{}\\
\llap{$^a$}University of Tokyo,\\
  3-1, Hongo 7-chome, Bunkyo-ku, Tokyo, Japan\\
\llap{$^b$}Brookhaven National Laboratory,\\
  Building 510, Upton NY 11973, US\\
\llap{$^c$}National Technical University Of Athens,\\
 Zografou Campus, 9, GR-159 73 Athens, Greece\\

  E-mail: \email{Konstantinos.Ntekas@cern.ch}}

\abstract{
Five small prototype micromegas detectors were positioned in the ATLAS detector during Large Hadron Collider running at $\sqrt{s} = 7$ and $8\, \mathrm{TeV}$. A $9\times 4.5\, \mathrm{cm^2}$ double drift gap detector was placed in front of the electromagnetic calorimeter and four $9\times 10\, \mathrm{cm^2}$ detectors on the ATLAS Small Wheel, the first station of the forward muon spectrometer. The one attached to the calorimeter was exposed to interaction rates of about $70\,\mathrm{kHz}/\mathrm{cm^2}$ at $\mathcal{L}=5\times 10^{33}\,\mathrm{cm^{-2}s^{-1}}$ two orders of magnitude higher than the rates in the Small Wheel.
We present the results from performance studies carried out using data collected with these detectors and we also compare the currents drawn by the detector installed in front of the electromagnetic calorimeter with the luminosity measurement in ATLAS.
}


\begin{document}

\section{Introduction}

\begin{figure}[H]
\centering
\includegraphics[scale=0.12]{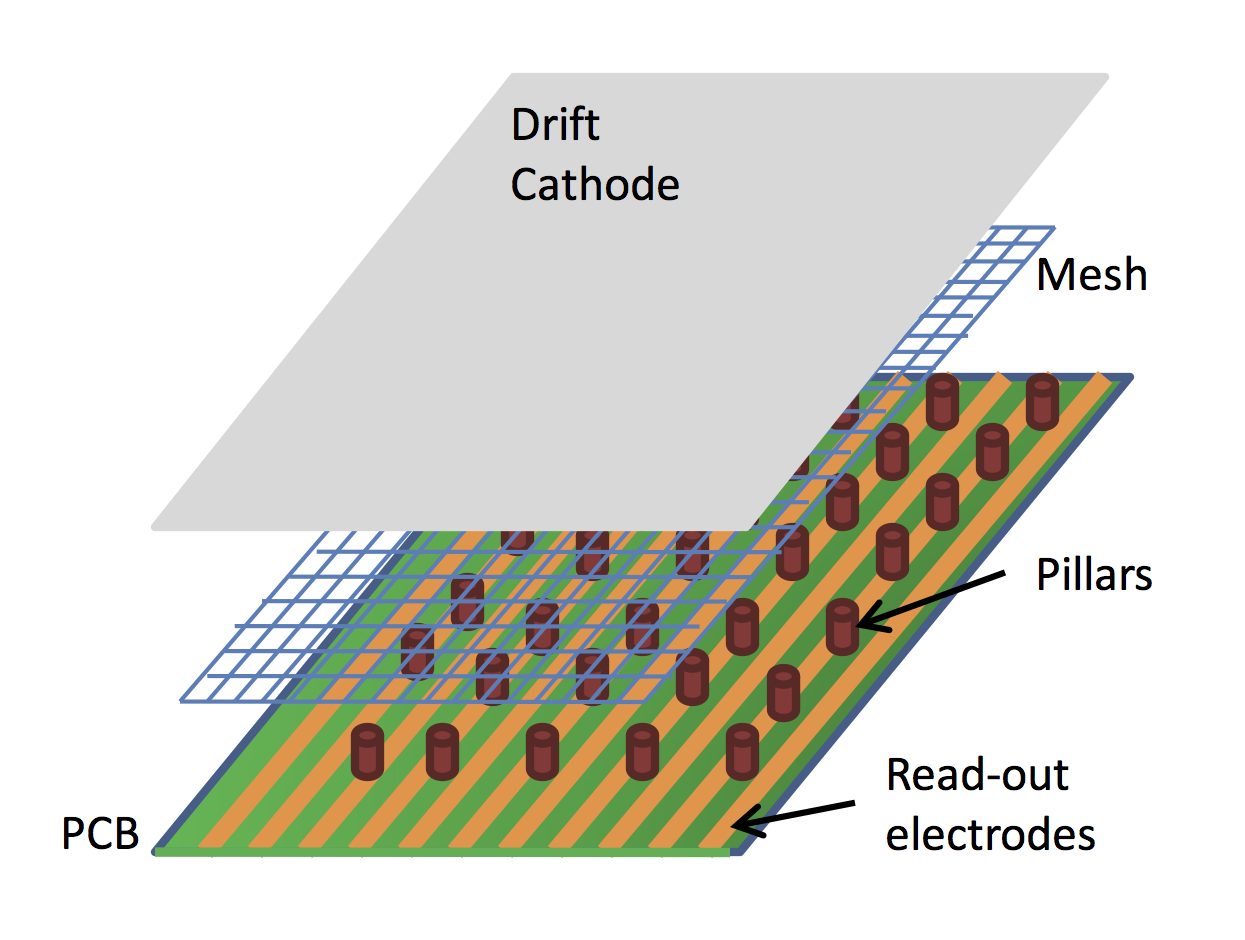}
\includegraphics[scale=0.28]{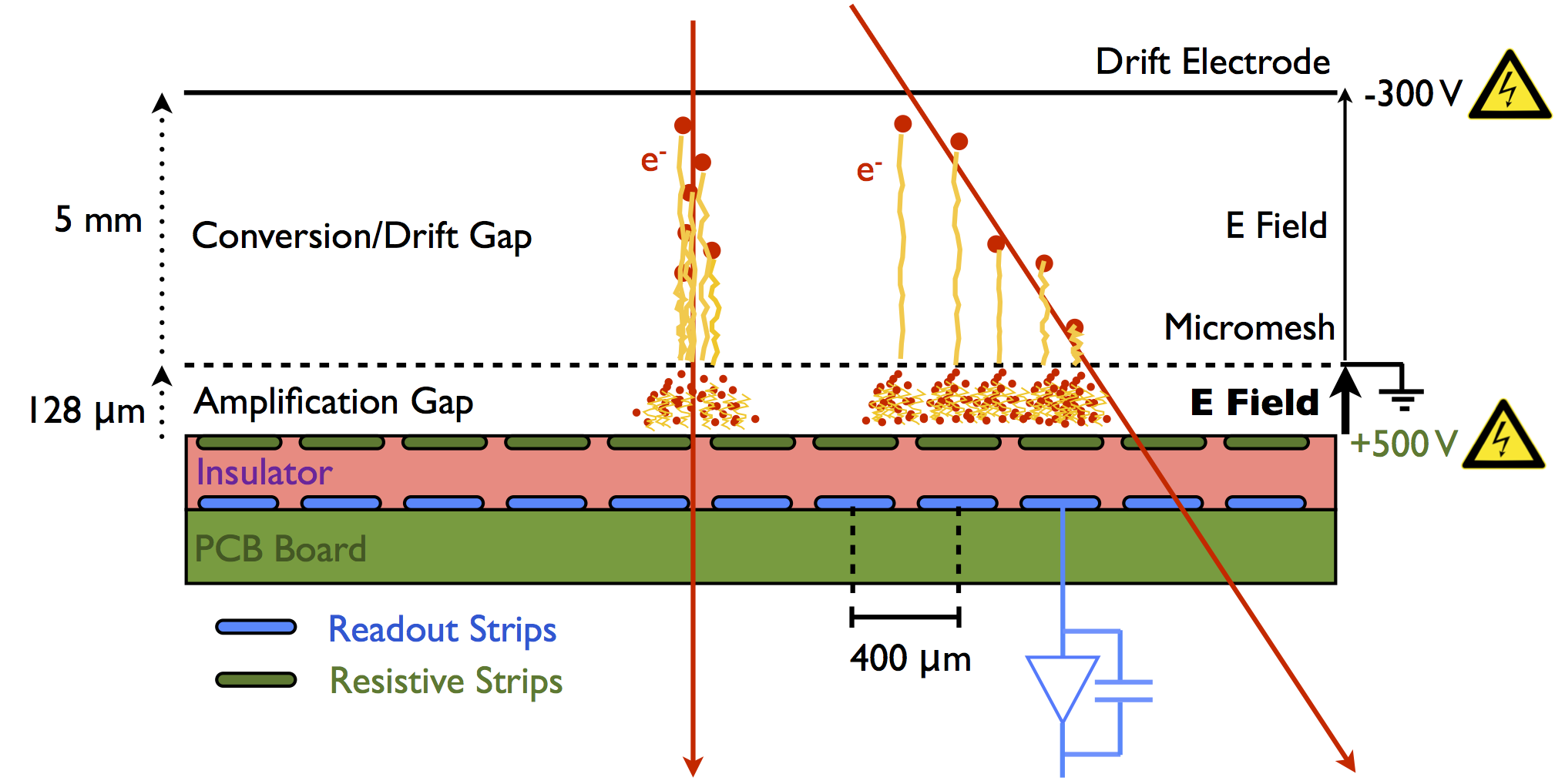}
\caption{Left: MM detector structure of a standard non resistive MM chamber Right: Principle of operation for a resistive MM chamber with $x$ readout strips of $400\,\upmu \mathrm{m}$ pitch.}  
\label{fig:mm_principle}
\end{figure}

The micromegas \footnote{An abbreviation for MICRO MEsh GAseous Structure.} technology was developed in the middle of the 1990's introducing a micropattern gaseous detector with two asymmetric regions \cite{micromegas_giomataris}. MM standard detectors consist of a planar (drift) electrode, a gas gap of a few millimeters thickness acting as conversion and drift region, and a thin metallic mesh at typically $100-150\,\upmu$m distance from the readout electrode, creating the amplification region, as shown in Figure \ref{fig:mm_principle}. The HV potentials are chosen such that the electric field in the drift region is a few hundred V/cm, and 40$-$50 kV/cm in the amplification region achieving gas gain values in the order of $\mathrm{10^4}$. \\

\noindent Charged particles traversing the drift space ionize the gas and the electrons liberated by the ionization process drift towards the mesh (tens of nanoseconds). With an electric field in the amplification region $50-100$ times stronger than the drift field, the mesh is transparent to more than $95\,\%$ of the electrons. The electron avalanche takes place in the thin amplification region in less than a nanosecond, resulting in a fast pulse of electrons on the readout strip \cite{ThirdReference}. The drifting of the electrons along with the avalanche formation in the amplification region for perpendicular and inclined tracks is illustrated in the right schema of Figure \ref{fig:mm_principle}. The ions that are produced in the avalanche process move towards to the mesh $\sim100$ times slower that the electrons. 

\begin{figure}[H]
\centering
\includegraphics[scale=0.7]{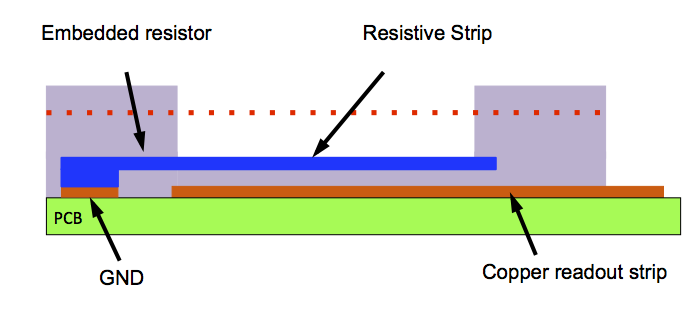}
\caption{Spark resistant micromegas. Sketch of the resistive and readout strip structure.} 
\label{fig:mm_res}
\end{figure}

\noindent Adding a layer of resistive strips on top of a thin insulator directly above the readout electrode, as shown in Figure \ref{fig:mm_res}, the MM become spark-insensitive. The readout electrode is no longer directly exposed to the charge created in the amplification region. The signals on the readout strips are partly induced signals and partly capacitively coupled to it. By adding this protection some fraction of the signal height is lost but the chamber can be operated at higher gas gain and thus have spark intensities reduced by about three orders of magnitude \cite{alexop}.\\

\noindent The MM technology is quite attractive for High Energy Particle Physics experiments due to its rate capability, excellent performance and
simplicity of construction. As a result an R$\&$D activity was started in 2007, called Muon ATLAS MicroMegas Activity (MAMMA), to explore the potential of the MM
technology for its use in LHC detectors. As of May 2013 ATLAS decided to equip the New
Small Wheel (NSW) with MM detectors \cite{NSW_TDR}, combining triggering and precision tracking functionality in a single device \cite{JoergReference}.

\section{Micromegas Detectors in ATLAS}

\noindent In order to test the micromegas detectors under realistic LHC conditions five small micromegas chambers were installed in the ATLAS detector in February 2012, until the end of the data taking in February 2013. One prototype resistive micromegas detector, with two gas gaps sharing a common drift plane (namely MBT03 and MBT04), with an active area of $9\times 4.5\,\mathrm{cm^2}$ was installed in the high-rate environment in front of the electromagnetic end-cap calorimeter (LAr), $3.5\,\mathrm{m}$ from the interaction point in the $z$ direction, at a radius $r\sim 1\,\mathrm{m}$. Four standard $9\times 10\,\mathrm{cm^2}$ resistive detectors (R13,16,18,19) were installed on the ATLAS Small Wheel (SW), Cathode Strip Chamber (CSC) region, at $1.8\,\mathrm{m}$ distance from the beam pipe.

\begin{figure}[H]
\centering
  \includegraphics[scale=.55]{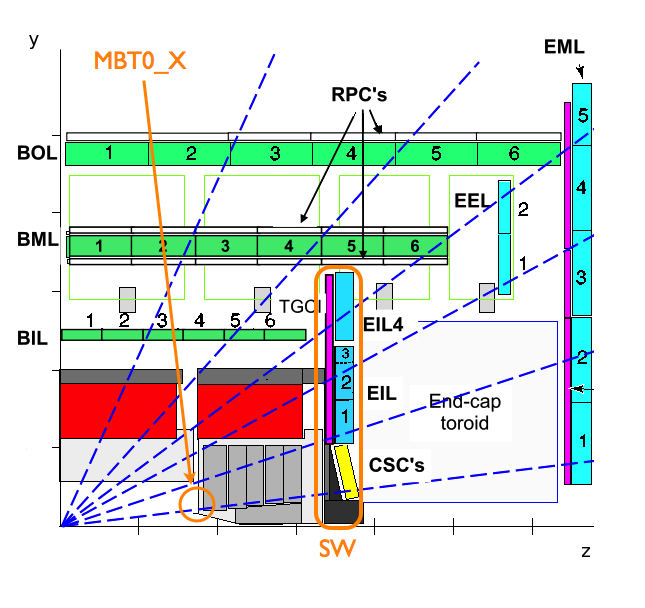}
  \caption{Sketch of the ATLAS detector. The place where the MBT chamber was installed is indicated with orange arrow while the SW region is also marked on the drawing.}
  \label{fig:MBT_sketch}
\end{figure}

\noindent In the work presented here we measure the hit rate of the different ATLAS regions where the micromegas chambers were installed. We demonstrate also, even with a random trigger, that recorded hits in the adjacent chambers can be correlated indicating real particle tracks. We then study the correlation of the MBT chamber's current with the luminosity of ATLAS demonstrating the capability of micromegas detector for luminosity measurement.

\subsection*{Chamber Characteristics}

\begin{figure}[H]
\centering
  \includegraphics[scale=.15]{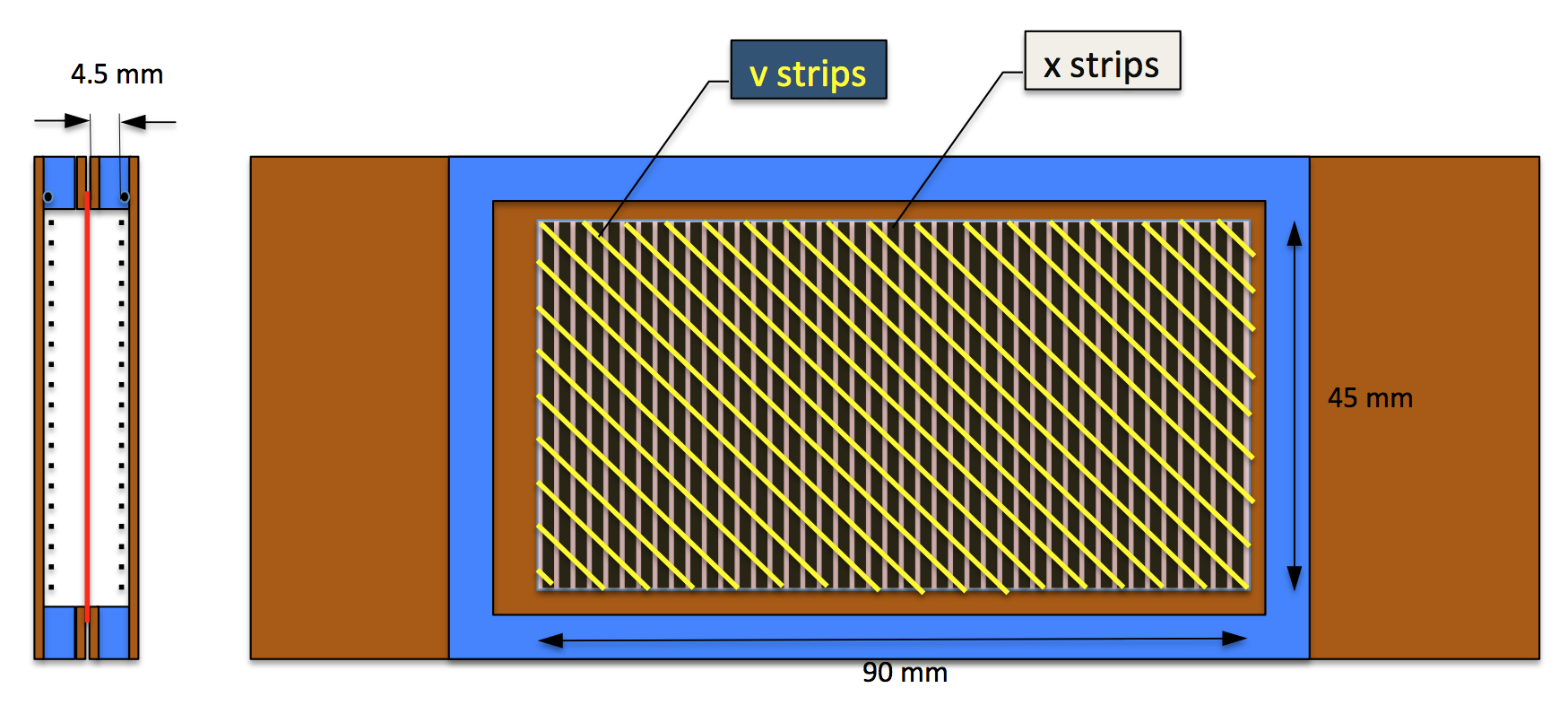}
  \caption{Sketch of the MBT double gap and strip configuration.}
  \label{fig:MBT_strips}
\end{figure}

The MBT is a $4.5\,\mathrm{mm}$ drift double-gap chamber, with an amplification gap of $0.128\,\mathrm{mm}$ and two-dimensional readout with $500\, \mathrm{\upmu m}$ pitch, ($400\, \mathrm{\upmu m}$ width) of the x strips (in $\upeta$-direction) and $1.3\,\mathrm{mm}$ pitch, ($200\,\mathrm{\upmu m}$ width) of the v strips which are inclined with respect to the x strips by $30^\circ$ (figure \ref{fig:MBT_strips}). It is a resistive-strip type micromegas with a resistivity of about $300\,\mathrm{M\Omega/cm}$ and a resistance of $100\,\mathrm{M\Omega}$. For the chamber operation we use a drift high voltage at $-300\,\mathrm{V}$ ($0.67\,\mathrm{kV/cm}$ nominal electric drift field) and amplification high voltage applied to the resistive strips of $+500\,\mathrm{V}$ ($39\,\mathrm{kV/cm}$ electric amplification field) while the mesh was connected to ground.\\

\noindent The micromegas detectors installed in the SW region are standard $9\times 10\,\mathrm{cm}^2$ resistive chambers with a $5\,\mathrm{mm}$ drift gap and two dimensional readout consisting of x and y strips $250\, \mathrm{\upmu m}$ pitch ($150\, \mathrm{\upmu m}$ width). The amplification gap is the standard $0.128\,\mathrm{mm}$. The resistive strips are $150\, \mathrm{\upmu m}$ wide with a resistivity of $200\,\mathrm{M\Omega/cm}$. The R chambers are operated with about the same high voltage configuration as the MBT with a bit higher voltage applied on the resistive strips to achieve the nominal drift electric field.\\

\noindent All chambers were operated with an $\mathrm{Ar}$:$\mathrm{CO_2}$ $(93$:$7)$ gas mixture and read out, stand-alone, by APV25\cite{APVReference} hybrid cards through the Scalable Readout System\cite{SRS}. They were continuously
operational except for a few weeks with low luminosity running. The high voltage was kept
on the same value independent of the LHC beam conditions (early beam, single bunches, large emittance beam, $25\,\mathrm{ns}$ LHC beams, etc). The detector currents were continuously monitored through the CAEN SY1527 high voltage system using the A1821 module with a monitor current resolution of $2\,\mathrm{nA}$\cite{caen1821}.

\begin{figure}[H]
	\centering
  \includegraphics[scale=.6]{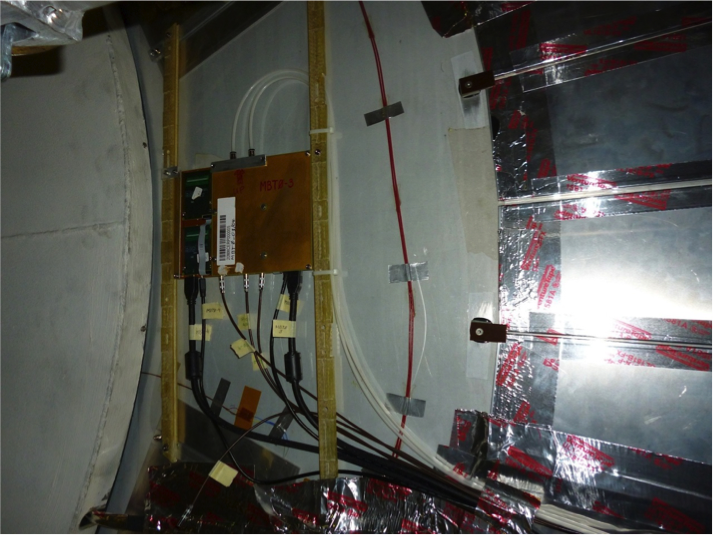}\hspace{15pt}
  \includegraphics[scale=0.077]{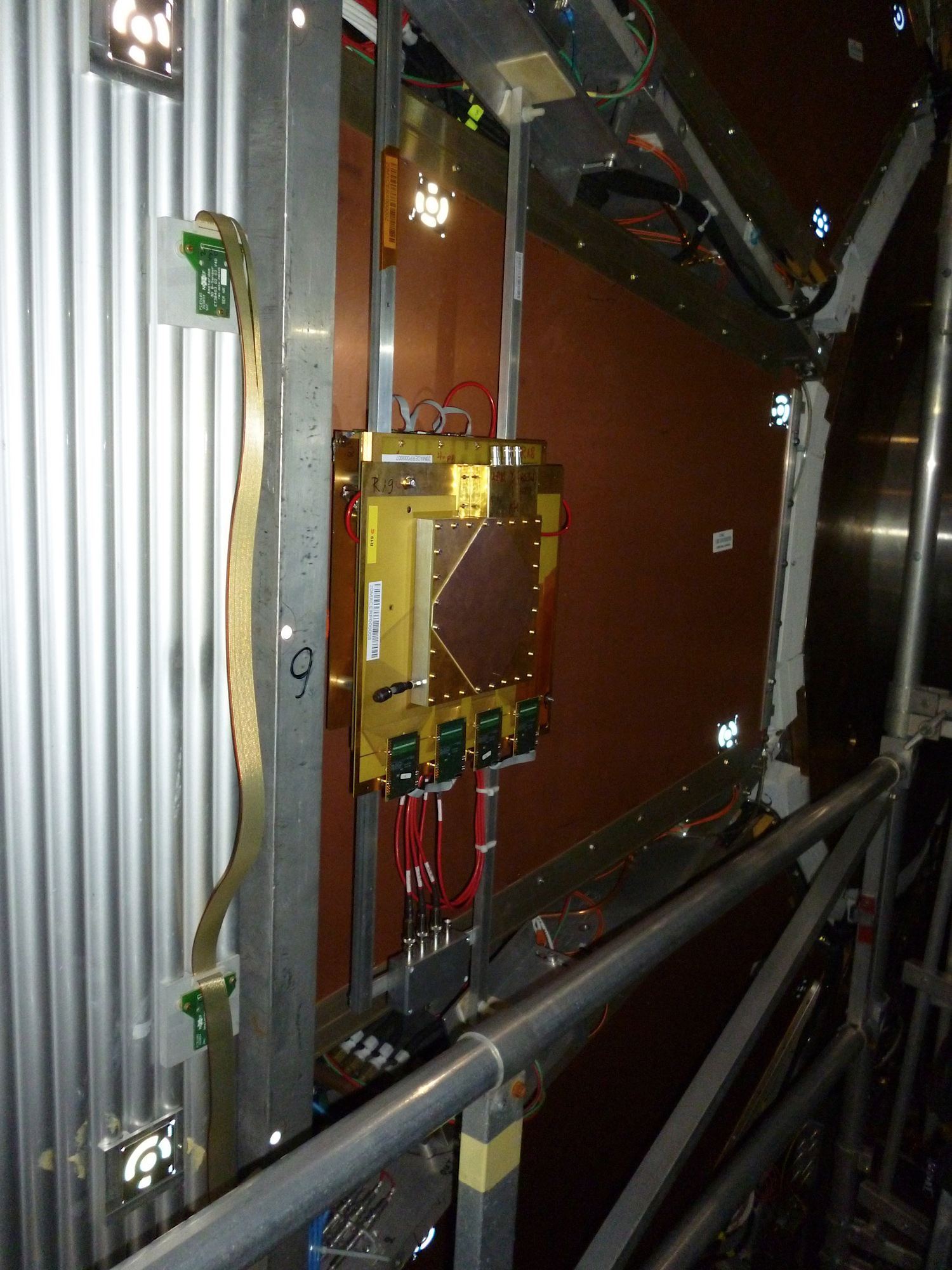}
  \caption{Left: MBT chamber installed on the LAr calorimeter. Right: R chamber installed on the ATLAS SW CSC region.}
  \label{fig:MBT_photo}
\end{figure}

\section{Hit Rate Measurement}

The luminous ATLAS experiment is a relatively new environment of operation for the micromegas detector. Therefore it was challenging to measure the recorded hit rate in the two different regions with the installed chambers. For this study we first apply a fiducial time bin cut on our events keeping only strips with measured time between 6 and 15 time bins (each bin width is equal to $25\,\mathrm{ns}$). Since the data were taken using a random trigger this time window ensures that the signal shape is fully recorded, from its rise to its falling tail. The earlier time window ($\leq 5\,$time bins) is mainly dominated by events where the rising edge happens before the start of the charge sampling by the APV25 and therefore is cut out. For times greater than the window's high edge  ($\geq 16\,$time bins) the signal tail is almost totally cut off. In both cases, the charge is not correctly measured and the measured rate would be inaccurate.\\

\begin{figure}[H]
\centering
	\includegraphics[width=.49\linewidth]{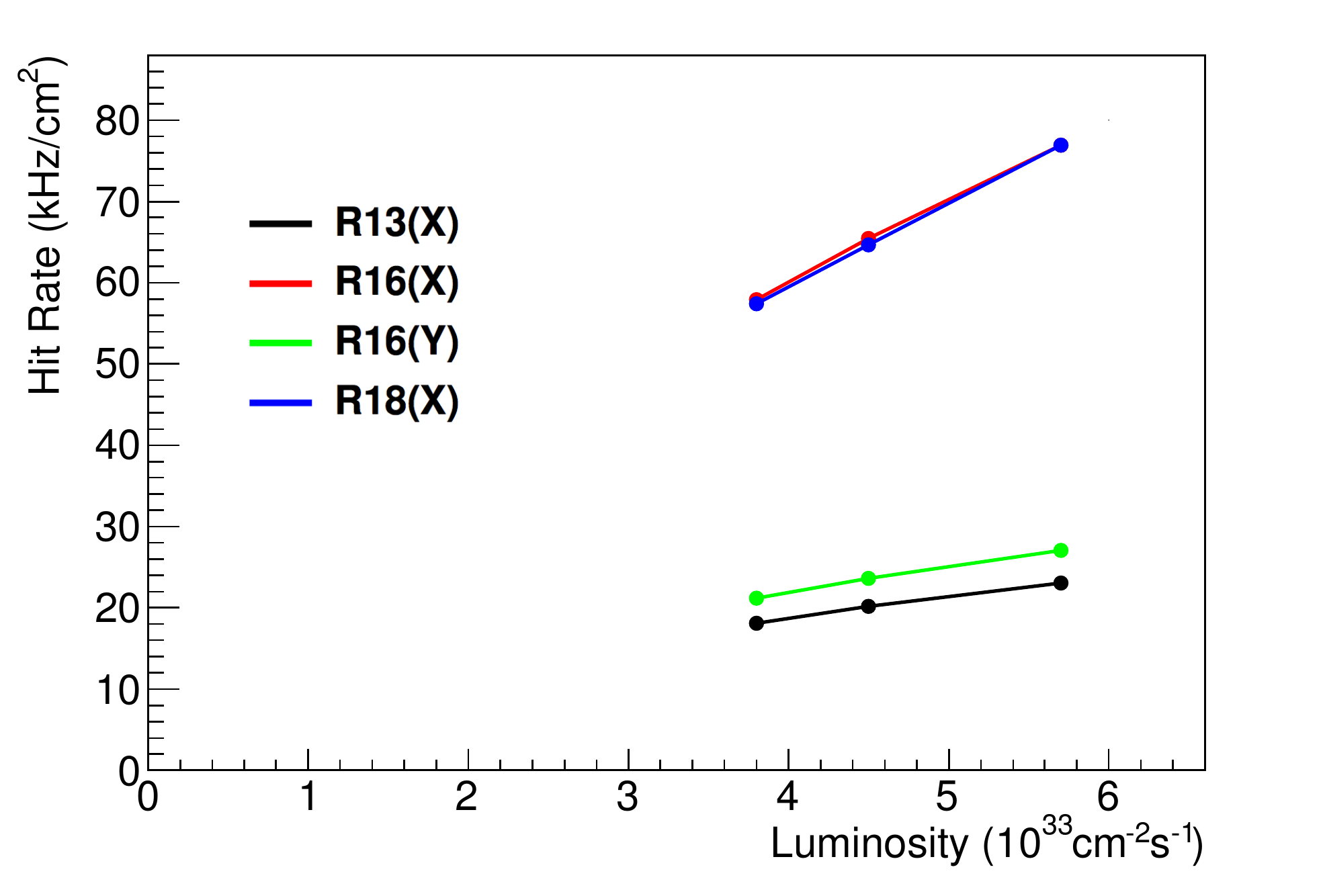}
	\includegraphics[width=.5\linewidth]{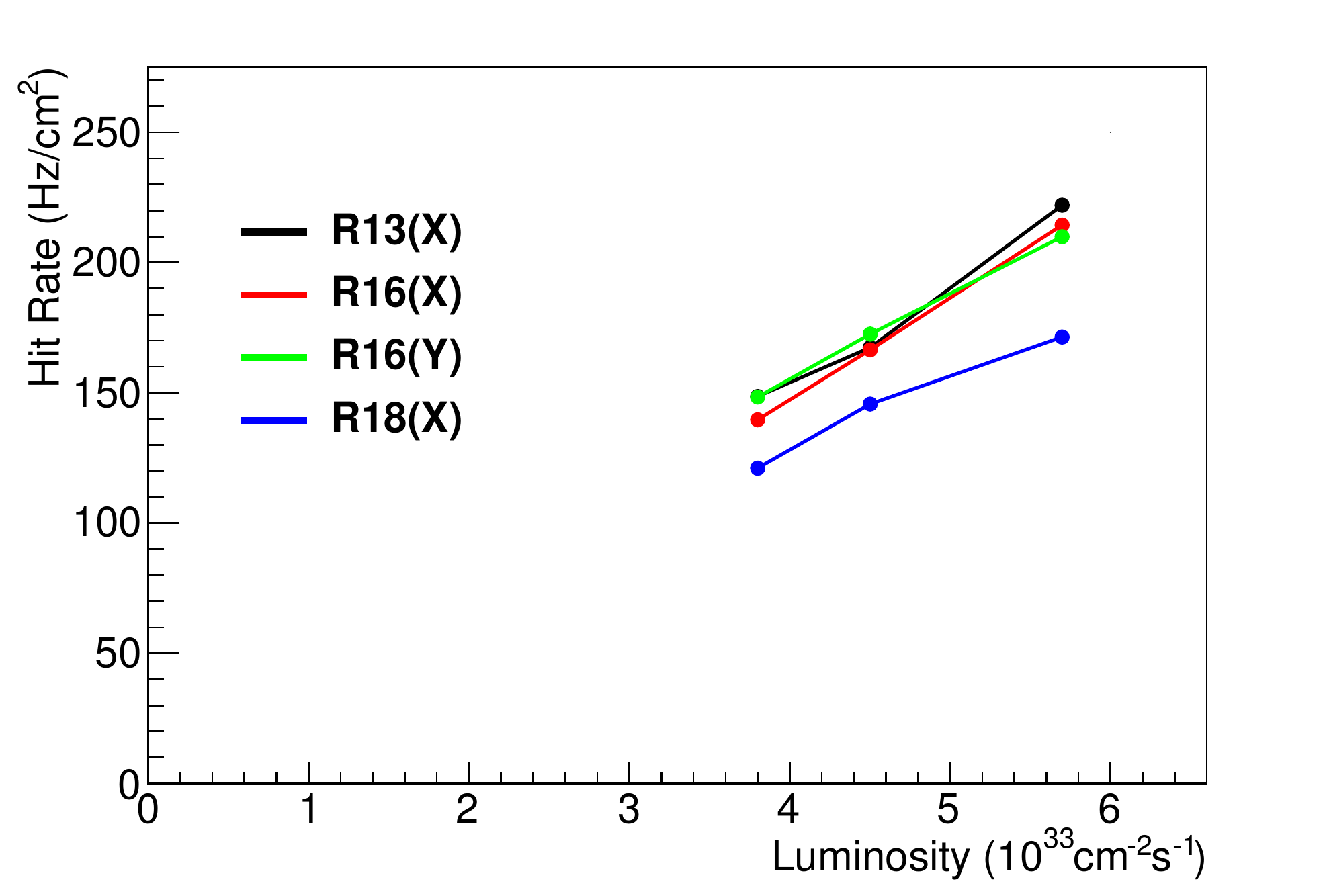}%
\caption{Measured hit rate as function of luminosity. Left: MBT chamber (2 gaps) in the LAr calorimeter region (red and blue points correspond to x readout strips while green and black points stand for the v strips measurement). Right: R chambers in the small wheel region.}	
\label{fig:hit_rate}
\end{figure}

\noindent We then apply a single clusterization algorithm to form clusters of strips. We allow up to 4 gaps between the strips that belong to the same cluster for the MBT chamber. For the small wheel chambers, due to the smaller strip pitch, we select a maximum gap of 10 strips for our clusterization. We then define the hit rate as the number of clusters counted per square centimeter per second and we present the measurements as a function of the ATLAS luminosity for the different chambers in figure \ref{fig:hit_rate}.

\noindent The hit rate of the small wheel chambers is measured in the order of $40\,\mathrm{Hz \, cm^{-2} \, s^{-1}}$ per $\mathcal{L}=10^{33}\,\mathrm{cm^{-2} \, s^{-1}}$  and almost proportional to the luminosity. This measurement should be compared with the $38\, \mathrm{Hz \, cm^{-2} \, s^{-1}}$ per $\mathcal{L}=10^{33}\,\mathrm{cm^{-2} \, s^{-1}}$ that is the value measured for CSC chambers in the same region. The MBT rate is about $15\, \mathrm{kHz \, cm^{-2} \, s^{-1}}$ per $\mathcal{L}=10^{33}\,\mathrm{cm^{-2} \, s^{-1}}$  and the extremely high hit rates deform the linearity a bit because the two hits are sometimes not separated in the clusterization. The v strips in MBT chambers show relatively low rates due to the wider clusters, due to the $30^\circ$ inclination with respect to the resistive strips, and strips with shorter length in some region.

\section{Track Identification}

\begin{figure}[H]
\centering
\includegraphics[width=0.495\linewidth]{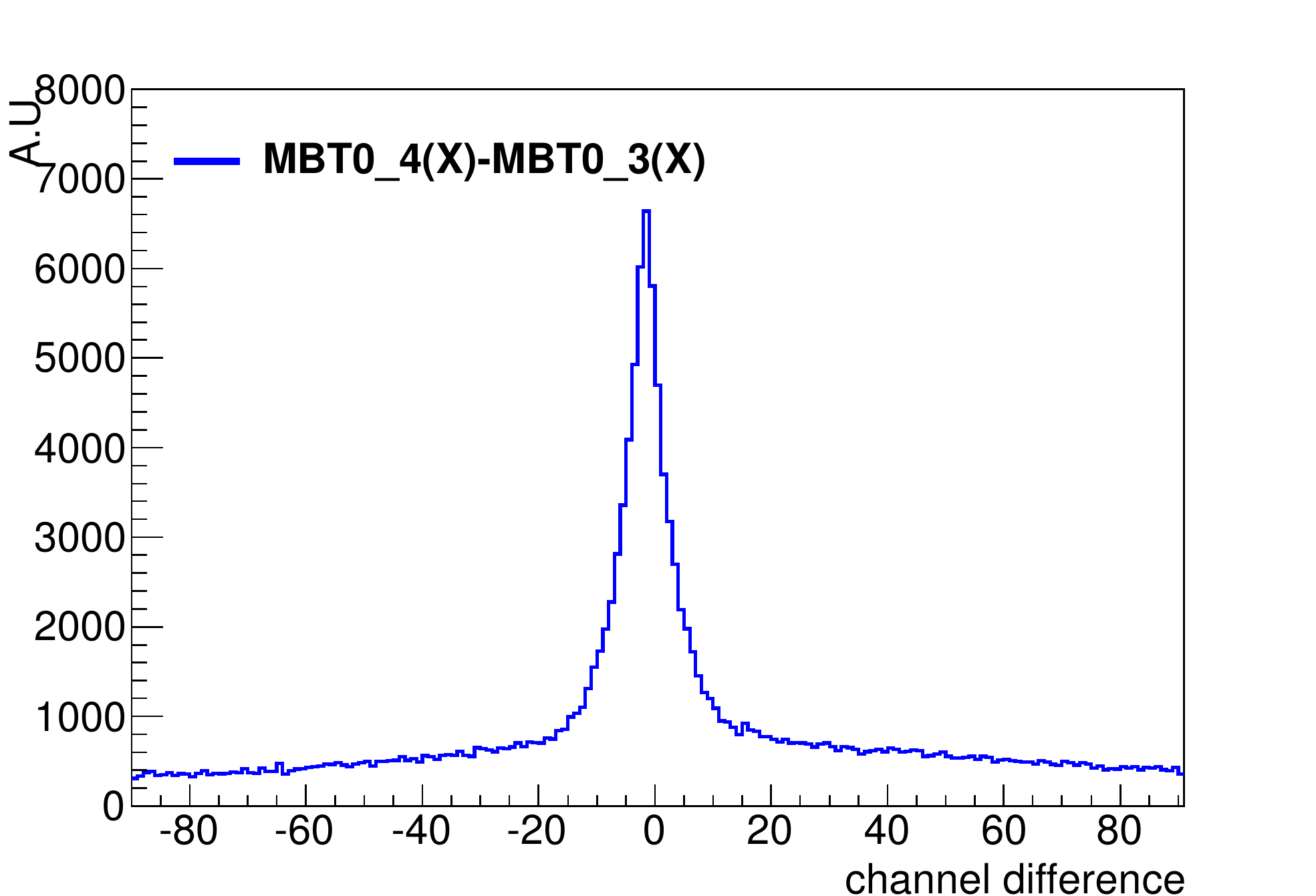}
\includegraphics[width=0.49\linewidth]{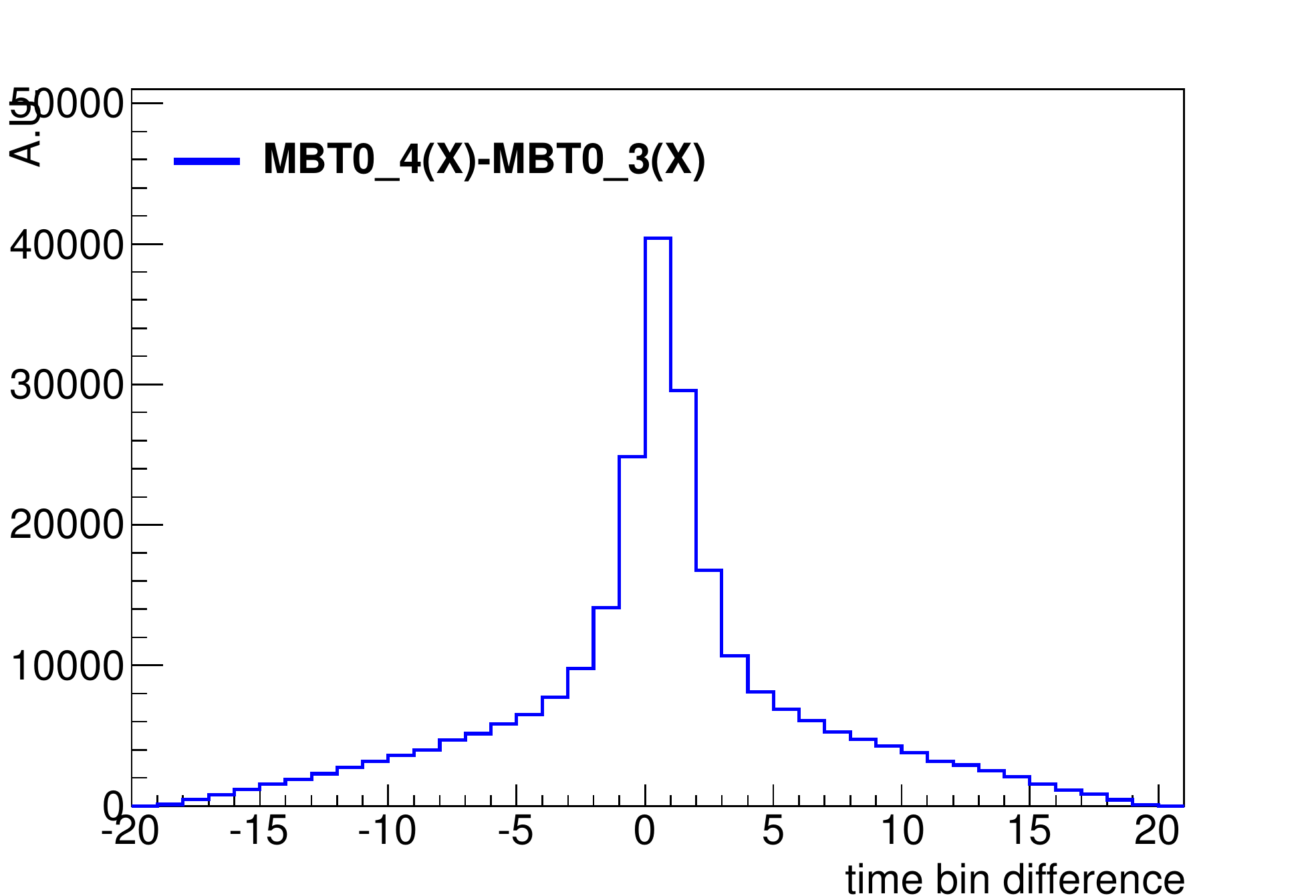}
\caption{Hit correlation for the consecutive gaps of the MBT chamber. Left: Hit position difference measured in channel number. Right: Hit time difference measured in time bin units (each time bin is $25\,\mathrm{ns}$ wide).}
\label{fig:track_id_mbt}
\end{figure}

\begin{figure}[H]
\centering
\includegraphics[width=0.49\linewidth]{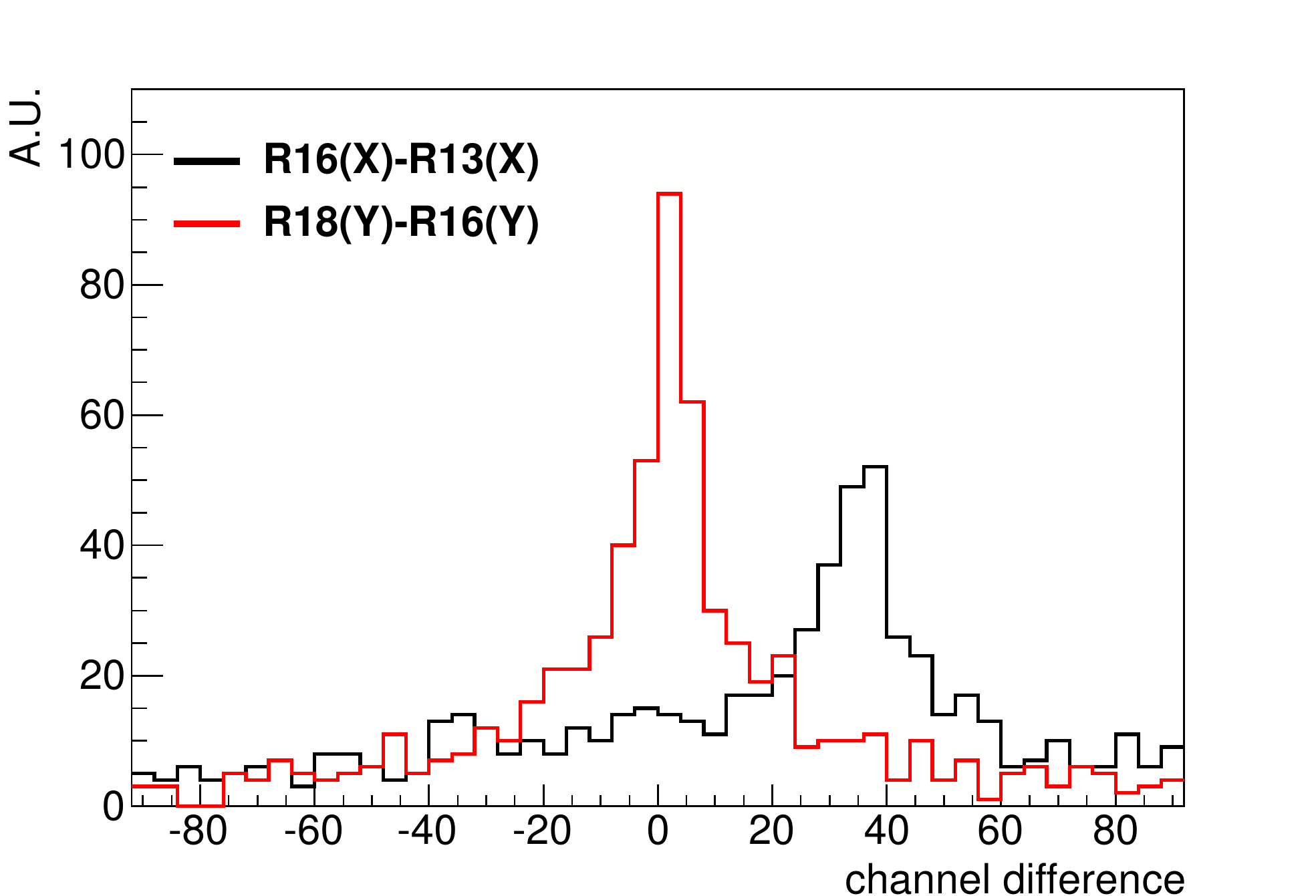}
\includegraphics[width=0.49\linewidth]{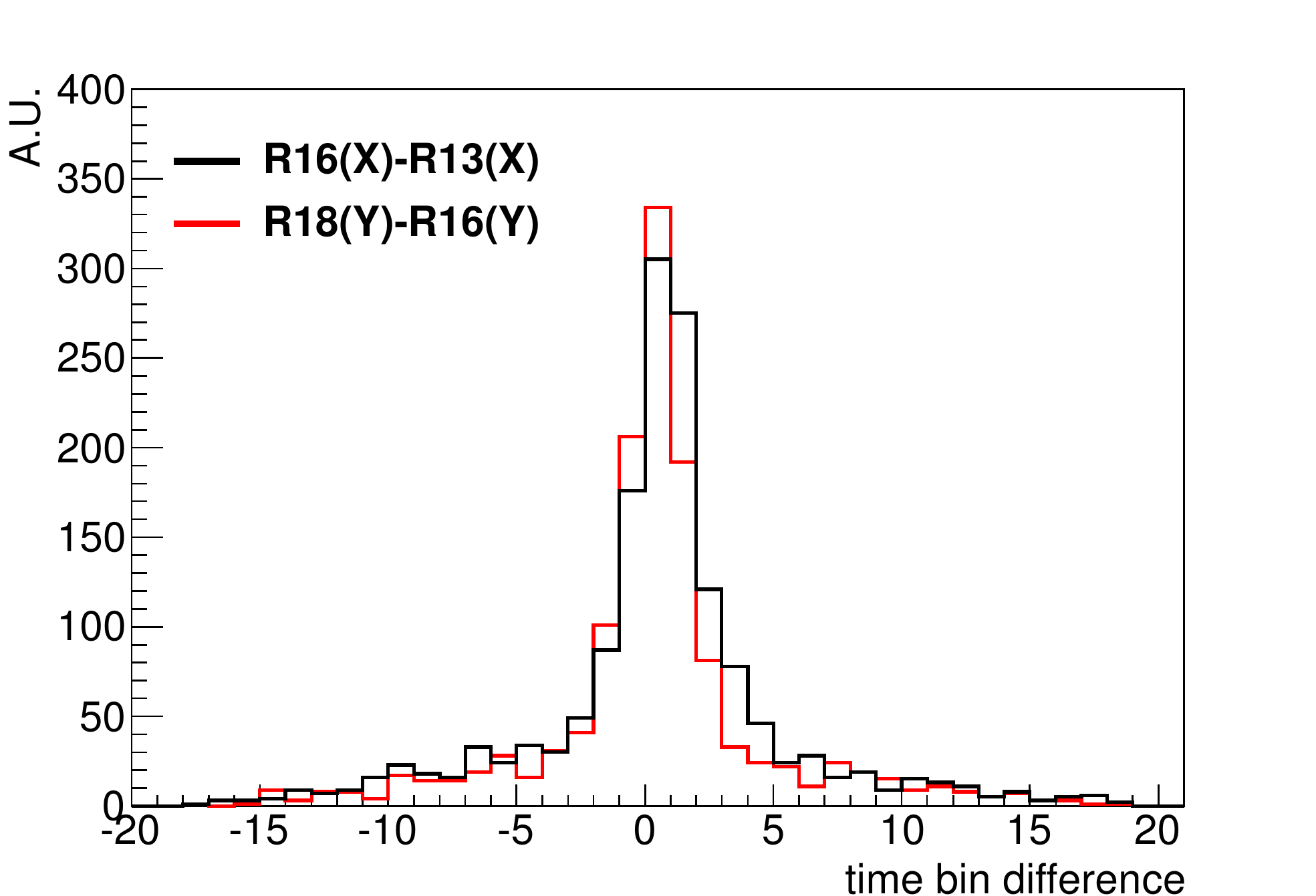}
\caption{Hit correlation for the consecutive R chambers. Left: Hit position difference measured in channel number. Right: Hit time difference measured in time bin units (each time bin is $25\,\mathrm{ns}$ wide).}
\label{fig:track_id_r}
\end{figure}

Correlated hits between chambers placed one after the other are considered as tracks passing through the chambers. These cases correspond to a $90\%$ fraction of events in MBT chamber, when comparing hits from MBT03 and MBT04 gaps, and about $10\%\sim 20\%$ in SW chambers. The difference of the measurement in the two regions emanates from the distance between the correlated chambers with the two gaps of the MBT having almost zero lever arm compared to the distances between the R chambers. The rest are treated as isolated hits from neutrons or gammas. Figures \ref{fig:track_id_mbt} and \ref{fig:track_id_r} illustrate the difference in position and time between two chambers, of the same area in the "real track events".
%\vspace{10pt}

\noindent The strong correlation is evident in both time and position plots for the two different areas, although the position difference depends on the incoming angle, especially when chambers have a large gap between them, which is the case of the SW chambers. The shift observed in the position difference for the x strips (measuring radius) in SW chambers, in the order of 35 channels, corresponds to the polar angle from the interaction point. The long tails of MBT chambers distributions emanate from combinatorial background due to the high rate, as shown in figure \ref{fig:track_id_mbt}.

\section{Luminosity Measurement}

\begin{figure}[H]
	\begin{center}
		\includegraphics[scale=.3]{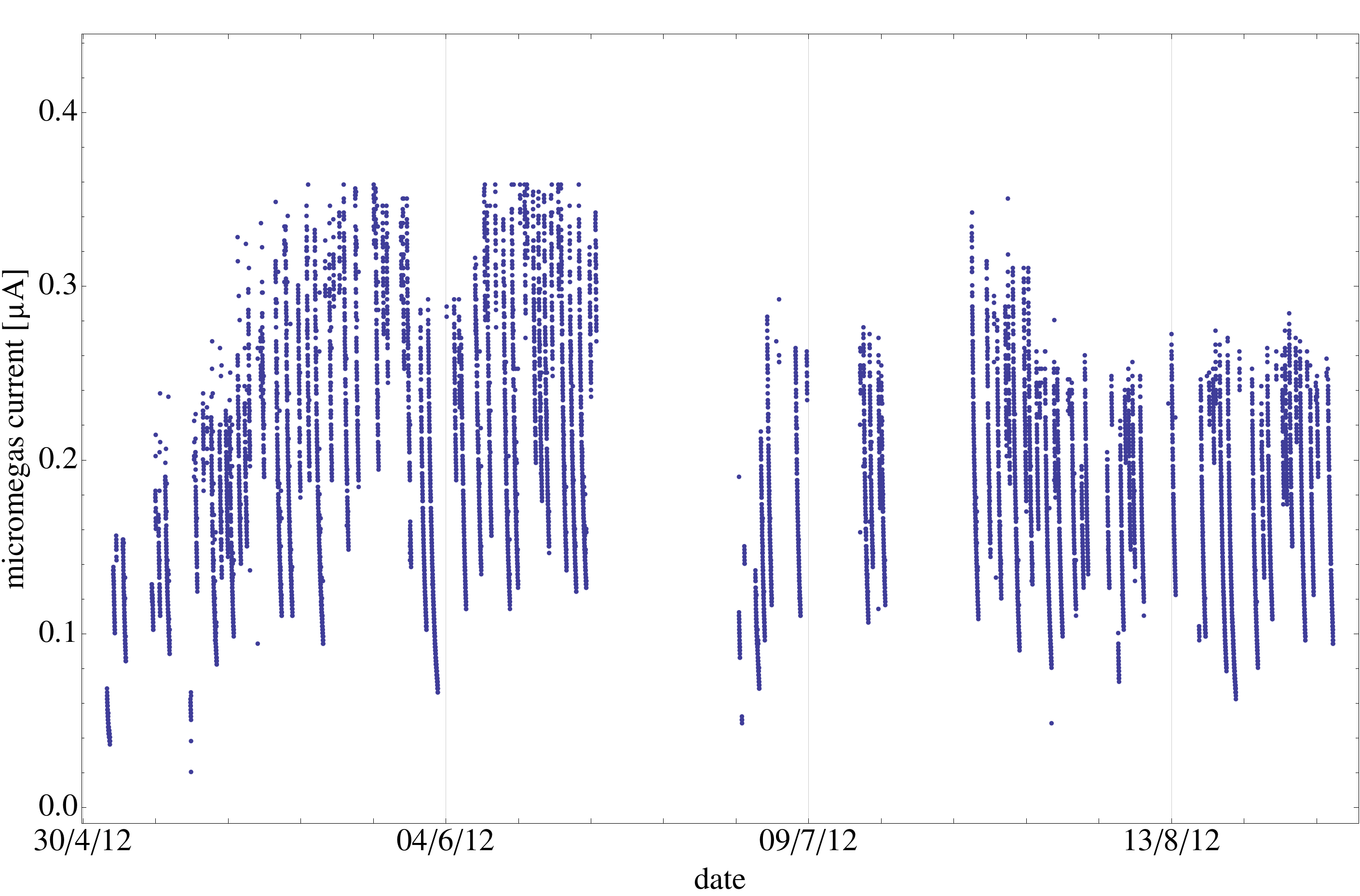}
		\caption{An example of micromegas currents as a function of time starting from early May until September 2012. The structure of the LHC fills is clearly visible.}
		\label{fig:PVSS}
	\end{center}
\end{figure}

As we have discussed in the introduction, our chambers where connected to a high voltage unit and we were monitoring their instantaneous current using PVSS (see figure \ref{fig:PVSS}) with a sensitivity of $2\,\mathrm{nA}$. The current drawn by the chambers should be a function of the incoming rate of particles and due to the high resolution of this detector technology a study has been performed trying to correlate the monitored instant current of the chambers with the instantaneous luminosity of the ATLAS experiment \cite{ATL_NOTE}.

\begin{figure}[H]\centering
		\includegraphics[width=.95\linewidth]{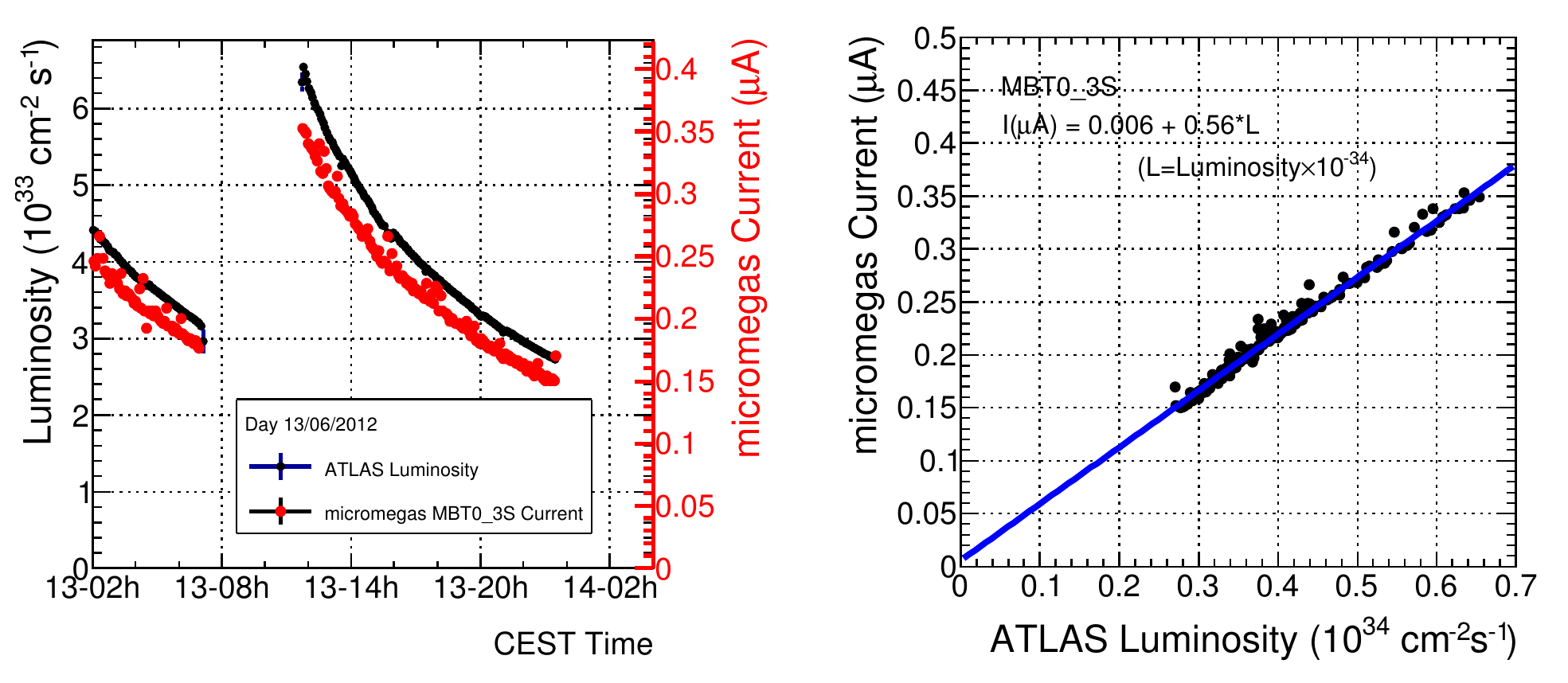}
		\caption{The left plot shows the ATLAS luminosity (left axis, black line) and the MBT current (right axis, red points) as a function of time for a given period. The right plot gives the MBT current versus the ATLAS luminosity for the same period. You can see a good correlation between the current and the luminosity. The blue line is a fit to the data.}
		\label{fig:day7}
\end{figure}

\noindent Correlation plots of the luminosity along with the micromegas MBT current, created using data recorded in one day, can be seen in figure \ref{fig:day7}. In the left plot, the micromegas current is plotted with the red points (measured in $\mathrm{\upmu A}$) and the ATLAS luminosity is plotted with black points as a function of time. 

\noindent The outlying current points correspond to sparks due to the high rate environment and should be excluded on a more detailed analysis. On the right plot the micromegas current is plotted versus the luminosity. The blue line is a first order polynomial fit performed on the datapoints. The fitted line's intercept is $6\, \mathrm{nA}$ and has a slope of $0.55\,\mathrm{\upmu A/10^{34}\, cm^{-2}s^{-1}}$. The spark points are visible in this plot as well and they do not follow the clear linear relation of the two measured values.\\

\begin{figure}[H]\centering
		\includegraphics[scale=.4]{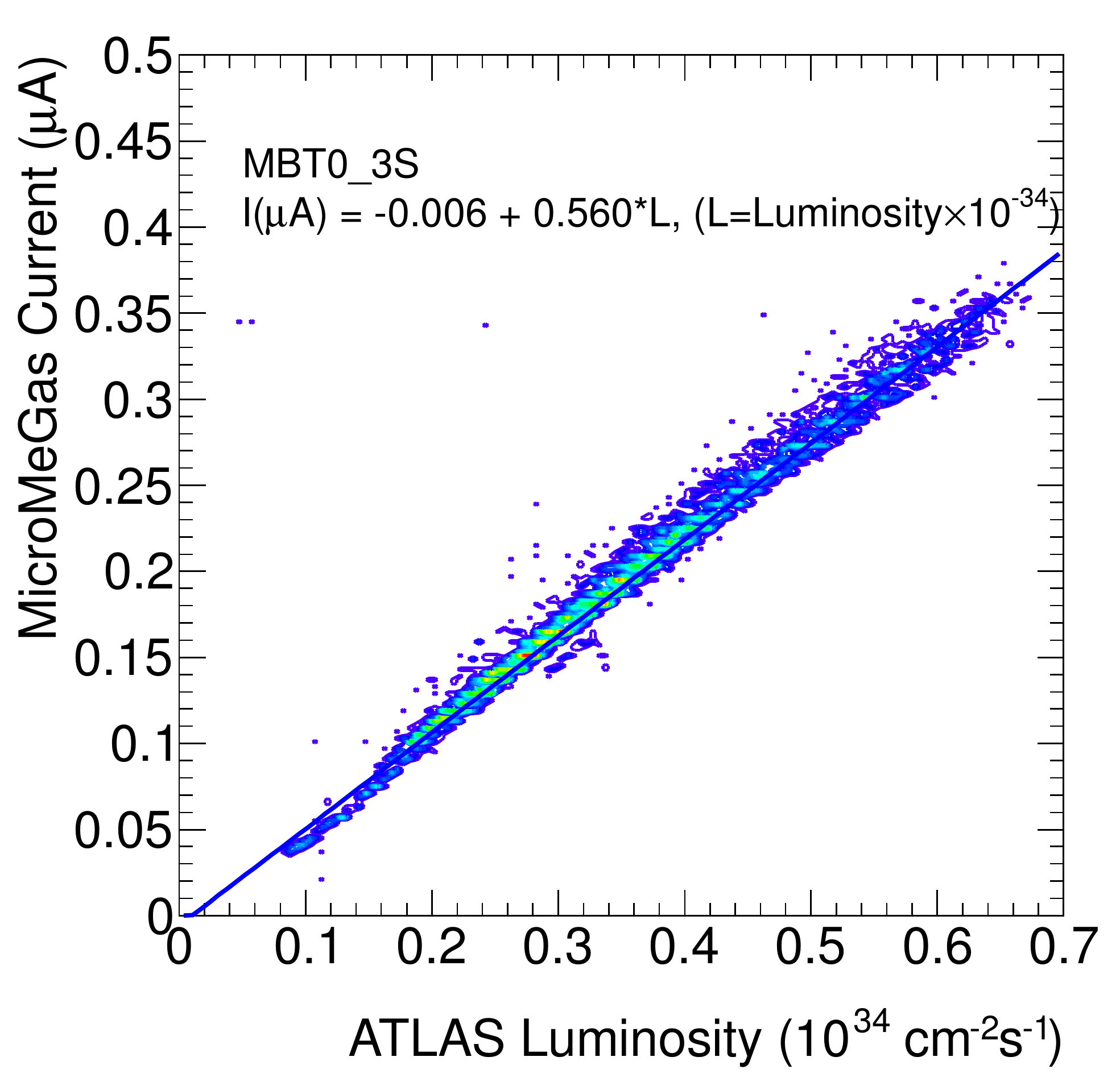}
		\caption{Correlation plot of the micromegas chamber and the luminosity as measured in the ATLAS experiment. The data are fitted with a first order polynomial (blue line).}
		\label{fig:all}
\end{figure}

\noindent Figure \ref{fig:all} shows the results of this analysis performed in the whole dataset. Performing a fit, we extract an intercept of $-6\ \mathrm{nA}$ and a slope of $0.56\ \mathrm{\upmu A/10^{34}\, cm^{-2}s^{-1}}$ that is in good agreement with the analysis of one days data shown in figure \ref{fig:day7}. A linear relation between the current of the micromegas chamber and the ATLAS luminosity was found (see figure \ref{fig:all}) and this allows the micromegas to be used for luminosity measurement along with LUCID and BCM\cite{BCM_LUCID}.

\noindent Using our measurement, we expect micromegas chamber of the size mentioned earlier to draw $0.56\, \mathrm{{\upmu} A}$ at the nominal ATLAS luminosity of $10^{34} \, \mathrm{cm^{-2}s^{-1}}$. Using this correlation, together with an integrated LHC luminosity of $15.5\,\mathrm{fb^{-1}}$, it is estimated that the total charge accumulated by the MBT chamber during the run adds up to about $1.2\,\mathrm{C}$ or $0.03\, \mathrm{C/cm^2}$. This figure should be compared to the accumulated charge of $0.005\, \mathrm{C/cm^2/year}$ that is expected in the micromegas operating at a gas gain of $10^4$ at a rate of $10\,\mathrm{kHz/cm^2}$, the highest counting rate expected on the SW. Or in other words, the one year long exposure of the MBT chamber located on the front face of the electromagnetic calorimeter is equivalent to six years of operating the micromegas on the NSW\cite{NSW_TDR} at $\mathcal{L}=5\times 10^{34}\,\mathrm{cm^{-2}s^{-1}}$.

\section{Conclusions}
We have successfully operated and readout micromegas detectors in a high-rate environment of the ATLAS cavern  for a year. Their voltages and current were constantly monitored without any problem even for the ones in front of the electromagnetic end-cap calorimeter ultra high-rate region. From the reconstructed hits that were recorded we were able to measure the hit rate with our chambers which in the case of the MBT was about $70\,\mathrm{kHz/cm^2}$ for a luminosity of $5\times 10^{33}\, \mathrm{cm^{-2}s^{-1}}$, with $90\%$ of the hits being correlated in both gaps.\\ 

\noindent We have also observed a strong linear correlation between the LHC luminosity and the current drawn by the micromegas detectors. This leads to the conclusion that it could be possible to use the micromegas detectors to monitor the LHC luminosity in the ATLAS cavern with an accuracy that needs to be measured. The measurement also shows that the micromegas chamber show no evidence of ageing effects as there is linear relationship between the current drawn and the ATLAS luminosity even for high luminosity values.

\acknowledgments
\noindent We would like to thank MAMMA collaboration and the people involved in the installation of the chamber setup in the ATLAS experiment providing the infrastructure for the high voltage control and monitoring and readout of the chambers. We would also like to thank Rui De Oliveira from the CERN/TE-MPE-EM group for his constant support and for the production of the micromegas detectors. Finally we would like to thank the reviewer for the valuable comments.\\

\noindent The present work was co-funded by the European Union (European Social Fund ESF) and Greek national funds through the Operational Program "Education and Lifelong Learning" of the National Strategic Reference Framework (NSRF) 2007-1013. ARISTEIA-1893-ATLAS MICROMEGAS.


\begin{thebibliography}{9}

\bibitem{micromegas_giomataris}Y. Giomataris et al., \emph{MICROMEGAS: a high-granularity position-sensitive gaseous detector for high particle-flux environments},
\href{http://www.sciencedirect.com/science/article/pii/0168900296001751}{\emph{Nucl. Instr. Meth.} \textbf{A 376} (1996) 29-35}.

\bibitem{ThirdReference}I. Giomataris et al., \emph{Micromegas in a bulk}, \href{http://www.sciencedirect.com/science/article/pii/S0168900205026501}{\emph{Nucl.Instr.Meth.} \textbf{A 560} (2006) 405-408}.

\bibitem{alexop}
  T. Alexopoulos et al.,
\emph{A spark-resistant bulk-micromegas chamber for high-rate applications},
\href{http://www.sciencedirect.com/science/article/pii/S0168900211005869}{
  \emph{Nucl. Instr. Meth.} 
  \textbf{A 640} (2011) 110-118}.

\bibitem{NSW_TDR}
ATLAS Collaboration, \emph{NSW TDR}, \href{http://cds.cern.ch/record/1552862?ln=en}{CERN-LHCC-2013-006. \textbf{ATLAS-TDR-020} (2013)}.
  
\bibitem{JoergReference}J. Wotschack, \emph{THE DEVELOPMENT OF LARGE-AREA MICROMEGAS DETECTORS FOR THE ATLAS UPGRADE}, \href{http://www.worldscientific.com/doi/pdf/10.1142/S0217732313400208}{\emph{Mod. Phys. Lett.} \textbf{A 28} (2013) 1340020}.

\bibitem{APVReference}M.J. French et al., \href{http://www.sciencedirect.com/science/article/pii/S0168900201005897}{\emph{Nucl. Instr. Meth.} \textbf{A 466} (2001) 359-365.}

\bibitem{SRS}
S. Martoiu et al.,
\emph{Development of the scalable readout system for micro-pattern gas detectors and other applications}, \jinst{8}{2013}{C03015}.
%JINST 8 C03015,
%(2013).

\bibitem{caen1821} CAEN, \emph{High Voltage power supply}, \textbf{A1821},  \href{http://www.caen.it/csite/CaenProd.jsp?parent=20&idmod=258}{\url{http://www.caen.it/csite/CaenProd.jsp?parent=20&idmod=258}}.

\bibitem{BCM_LUCID}
ATLAS collaboration,
\emph{Luminosity Determination in pp Collisions at $\sqrt{s}=7\,\mathrm{TeV}$ Using the ATLAS Detector at the LHC}, \href{http://arxiv.org/abs/1302.4393}{\emph{arXiv:\textbf{1302.4393}}}, (2013).

\bibitem{ATL_NOTE}
ATLAS collaboration,
\emph{Using Micromegas in ATLAS to Monitor the Luminosity}, \href{https://cds.cern.ch/record/1637033}{\emph{ATL-MUON-PUB-\textbf{2013-002}}}, (2013).

\end{thebibliography}
\end{document}